\documentclass[showkeys,twocolumn,showpacs,preprintnumbers,amsmath,amssymb]{revtex4}
\usepackage{graphicx}
\usepackage{dcolumn}
\usepackage{bm}
\begin{document}
%
%
\title{Dynamical systems as logic gates}
\author{Madhekar Suneel}
\affiliation{PGAD, Defence Research and Development Organization (Ministry of Defence, Government of India), DRDL Complex, Kanchanbagh, Hyderabad - 500 058, India.}
\date{\today}

\begin{abstract}
A scheme for logical computation using non-linear dynamical systems is presented.  Examples of discrete-time maps configured as AND, OR, NAND and NOR gates are given.  It is seen that the logical operations are \emph{flexible} in the sense that an AND gate can be transformed into an OR gate with a simple change of a single parameter and vice-versa.  Also, a NAND gate can be transformed into a NOR gate and vice-versa.  It is shown, by example, that the scheme can even be extended to continuous-time flows.  Since the NAND and NOR operations are universal, it is possible to implement any switching function by interconnecting blocks that realize these operations.

\keywords{Nonlinear systems, bifurcation, logic, computation}

\pacs{05.45.-a}

\end{abstract}

\maketitle

\section{Introduction}
In recent years, computation using non-conventional techniques has received considerable attention.  Computation and information-processing techniques based on chemical, biological and quantum-mechanical phenomena have been studied.  Murali, Sinha, Ditto and Munakata have proposed using clipped chaotic-dynamics for carrying out flexible computations \cite{munakata02, murali03, murali03_2}.  In this paper, a simple scheme is proposed to configure any non-linear dynamical system as a logic gate.  For this purpose, the qualitative change undergone by the system-dynamics at a bifurcation is exploited.  The principle of construction of the AND, OR, NAND and NOR gates is demonstrated by examples.  Both discrete-time and continuous-time nonlinear dynamical systems are considered.

\section{Computation using One-Dimensional Maps}
Consider the logistic map \cite{may76}
\begin{equation}
  x_{n+1} = \lambda \: x_{n} \: \left(1-x_{n}\right) \enspace .
\end{equation}
$\lambda$ is the control parameter, $n$ is the discrete-time and $x$ is the state of the system. Let $\lambda = {\lambda}' + \lambda_a + \lambda_b \enspace,$ where ${\lambda}'$ is a \emph{bias}.  $\lambda_a$ and $\lambda_b$ are logic inputs that take the value $L_0$ for logic-$0$ (false) and $L_1$ for logic-$1$ (true).

\begin{figure}[ht]
\begin{center}
\includegraphics[scale = 0.2]{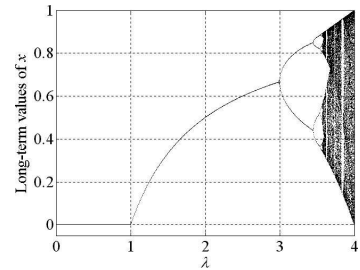}
\caption{Bifurcation diagram of the logistic map.}
\label{fig:bifurcation_logistic}
\end{center}
\end{figure}

For the logistic map, the first period-doubling bifurcation takes place at $\lambda \approx 3$.  The second period doubling takes place at $\lambda \approx 3.45$.  The bifurcation diagram of the logistic map is shown in Fig. \ref{fig:bifurcation_logistic}.

Let $L_0=0$ and $L_1=0.29$ in conformance with positive logic.  Let us choose ${\lambda}' = 2.56$.  Then, if both $\lambda_a = \lambda_b = L_0$, $\lambda = {\lambda}' = 2.56$ and the map will converge to a fixed point.  If either $\lambda_a$ or $\lambda_b$ is equal to $L_1$, $\lambda = {\lambda}' + L_1 = 2.85$ and the map still converges to a fixed point.  However, if both $\lambda_a = \lambda_b = L_1$, $\lambda = 3.14$ and the map settles into a 2-cycle.  If the logic output $Y$ is regarded as a False (logic 0) for a fixed point and a True (logic 1) for a 2-cycle, then it can be seen that the map functions as an AND gate.

If the value of ${\lambda}'$ is increased to $2.85$ while retaining the values of $L_0$ and $L_1$ , the system settles into a 2-cycle if either or both of $\lambda_a$ and $\lambda_b$ are true and hence we obtain an OR gate.  Therefore, it can be seen that a suitable change in the parameter ${\lambda}'$ transforms the OR gate into an AND gate and vice-versa. The scheme is summarized as Table \ref{tab:ParameterCumTruthTable1}.
\begin{table}[ht]
\caption{Logistic map as an AND/OR Gate.  $Y=0$ for a fixed point, $Y=1$ for a 2-cycle.  The transfer-function is controlled by the bias ${\lambda}'$.}
		\begin{center} 
		\begin{ruledtabular}
		\begin{tabular}{lllllll}
			$\left(L_0,L_1\right)$ & ${\lambda}'$ & $\lambda_a$ & $\lambda_b$ & $\lambda$ & $Y$ & Gate \\ \hline
		~ 										& ~ 		  		& 0 & 0 & $2.56$ & 0 & ~ 		\\
		~ 										& ~ 		  		& 0 & 1 & $2.85$ & 0 & ~ 		\\
		~ 										& 2.56 			& 1 & 0 & $2.85$ & 0 & AND 	\\
		~ 										& ~ 		  		& 1 & 1 & $3.14$ & 1 & ~ 		\\
		\multicolumn{7}{l}{$\left(0,0.29\right)$ \hrulefill}  \\
		~ 										& ~ 		  		& 0 & 0 & $2.85$ & 0 & ~ 		\\
		~ 										& ~	  		& 0 & 1 & $3.14$ & 1 & ~			 	\\
		~ 										& 2.85			 			& 1 & 0 & $3.14$ & 1 & OR 		\\
		~ 										& ~ 		  		& 1 & 1 & $3.43$ & 1 & ~ 			
		\end{tabular}
		\end{ruledtabular}
	\end{center}
	\label{tab:ParameterCumTruthTable1}
\end{table}

\begin{figure}[ht]
        \includegraphics[scale = 0.325]{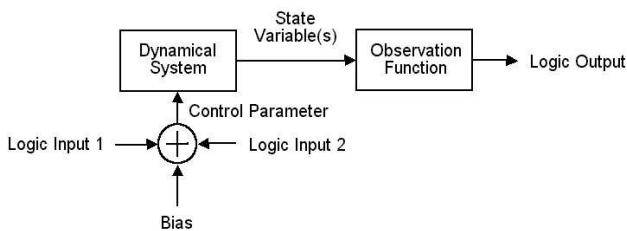}
\caption{Implementation of a logic-gate using a dynamical system.  The transfer-function of the gate can be modified using the bias-input.  The observation function transforms the state of the system into a \emph{conventional} logic output.}
\label{fig:scheme1}
\end{figure}

In this scheme, the logic output $Y$ is defined based on the state of the system.  Such a definition of the output is not in line with the output of \emph{conventional} (electrical) logic gates that have two-level output signals.  However, it must be mentioned that definition of input/output logic based on observed phenomena in the state of the system is not entirely new.  For instance, Steinbock et al have proposed realizations of chemical-wave logic circuits wherein the binary output logic is defined by the synchronous or asynchronous nature of waves propagating through a medium \cite{steinbock96}.  For compatibility with conventional logic gates having two-level inputs and outputs, one can define an \emph{Observation Function} that transforms the state of the dynamical system into a two-level signal.  Note that the observation function is not essential for the system to function as a logic gate.  However, if it is required to inter-connect logic gates, two-level inputs and outputs are helpful.  Realizable observation functions with an appropriate level-shifting mechanism should make it possible to connect the logic output of one gate as a logic input for another.  For the current example of the logistic map, an observation function $F$ can be defined to give meaningful logic output as
\begin{equation}
  F\left(x,n\right) = \left|x_n - x_{n-1}\right| \enspace .
\end{equation}
A fixed-point, with this observation function, yields zero whereas a 2-cycle yields the distance between the two points of the attractor set, which is a positive real number.  Therefore, observing the output of this function is equivalent to assuming \emph{a fixed-point denotes logic 0 and a 2-cycle denotes logic 1}.  The scheme of implementation of a logic gate is summarized in Fig. \ref{fig:scheme1}.

In the preceding example, an AND/OR gate is constructed around the period-doubling from a fixed-point to a 2-cycle.  In principle, the scheme should work around any bifurcation of the map.  Since the behavior of the system undergoes a noticeable qualitative change at the point of bifurcation, if an observation function that differentiates between the two behaviors can be defined, the system can function as a logic gate.  The bias of the control parameter (${\lambda}'$  in the preceding example) and the values of the logical inputs $L_0$ and $L_1$ can be chosen in several ways.  However, the author has used the following scheme for this choice:  An interval $\left(U,V\right)$ is chosen such that the critical value of the control parameter $\lambda = \lambda_c$ is exactly at its middle.  $U$ and $V$ should respectively be chosen such that the map demonstrates the desired pre-bifurcation and post-bifurcation behaviors at these points.  In the case of the preceding example, $\lambda_c \approx 3$ and the map undergoes a further bifurcation to a 4-cycle if $\lambda > 3.44$, a distance of about $0.44$ from $\lambda_c$.  Therefore, we choose the interval $\left(U,V\right) = \left(\lambda_c - 0.44, \lambda_c + 0.44 \right)$ i.e. $\left(2.56,3.44\right)$.  We find the value $w = \left(V-U\right)/3 \approx 0.29$.  Then we take $L_0 = 0$ and $L_1 = w$.  For operation as an AND gate, we choose ${\lambda}' = U$ and for operation as an OR gate, we choose ${\lambda}' = U+w$.

\section{Computation using Two-Dimensional Maps}
As a second example of computation using discrete-time maps, consider the Duffing's map:
\begin{equation}
\label{eqn:Duffings}
	\left .
    \begin{array}{lcl}
      x_{n+1} & = & y_n \\
      y_{n+1} & = & -\beta x_n + \alpha y_n - {y_n}^3  \enspace \enspace
    \end{array}
  \right\}
\end{equation}

\begin{figure}[ht]
      \includegraphics[scale = 0.185]{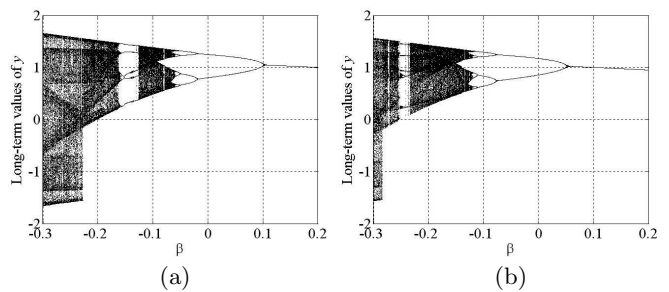}
    \begin{tabular*}{\linewidth}{cc}
    \makebox[0.5\linewidth][c]{(a)} & \makebox[0.5\linewidth][c]{(b)}
  \end{tabular*}
\caption{Bifurcation diagrams of the Duffing's map.  In (a), $\alpha = 2.2$.  In (b), $\alpha = 2.1$.  When $\alpha$ is reduced from $2.2$ to $2.1$, the critical point of bifurcation from a fixed-point to a 2-cycle changes from $\beta \approx 0.1$ to $\beta \approx 0.05$.  Such changes in the critical point can be exploited to control the transfer-function of the logic gate (see text for details).}
\label{fig:duffingbif}
\end{figure}

The bifurcation diagrams of Eq. (\ref{eqn:Duffings}) at $\alpha=2.2$ and $\alpha = 2.1$ respectively are shown in Fig.s \ref{fig:duffingbif}(a) and \ref{fig:duffingbif}(b).

It can be noticed that the bifurcation diagrams are qualitatively different from that of the logistic map since progressive bifurcations take place with progressively \emph{decreasing} values of the control parameter $\beta$.  This fact makes it possible to configure the Duffing's map as a NAND/NOR gate instead of AND/OR.  Analogous to the previous example, let $\beta = {\beta}' + \beta_a + \beta_b$.  First, let us work with $\alpha = 2.2$.  In this case, a bifurcation from a fixed-point to a 2-cycle takes place at $\beta \approx 0.1$. Let $L_0 = 0$ and $L_1 = 0.066$.  As in the case of the logistic map, let a fixed-point denote false ($Y=0$) and a 2-cycle denote true ($Y=1$).  Then a NAND gate and a NOR gate can respectively be realized with  ${\beta}' = 0$ and ${\beta}' = 0.066$.

Looking at Fig.s \ref{fig:duffingbif}(a) and \ref{fig:duffingbif}(b), one more way of transforming the NAND gate realized at $\left(\alpha = 2.2, {\beta}'=0\right)$ to a NOR gate becomes apparent.  Instead of changing the value of ${\beta}'$, if the value of $\alpha$ is changed to $2.1$, the bifurcation from a fixed-point to a 2-cycle occurs at  $\beta \approx 0.05$.  If we retain $\left({\beta}'=0, L_0 = 0, L_1 = 0.066\right)$, the system settles to a fixed-point if at least one of $\beta_a, \beta_b$ equal $L_1$ and the system functions as a NOR gate.  Since $\beta$ is considered the control parameter, we refer to $\alpha$ as the \emph{secondary} control parameter.  Transfer-function control of the logic gate by bias value as well as secondary control parameter is summarized in Table \ref{tab:ParameterCumTruthTable2}.
\begin{table}[ht]
\caption{Duffing's map as a NAND/NOR gate.  $Y=0$ for a fixed point, $Y=1$ for a 2-cycle.  The transfer-function can be controlled by bias ${\beta}'$ or secondary control parameter $\alpha$.}
		\begin{center} 
		\begin{ruledtabular}
		\begin{tabular}{lllllll}
			$\left(L_0,L_1\right)$ & $\left(\alpha,{\beta}'\right)$ & $\beta_a$ & $\beta_b$ & $\beta$ & $Y$ & Gate \\ \hline
		~ 										& ~ 		  								& 0 & 0 & $0$ 		& 1 & ~			\\
		~ 										& ~ 		  								& 0 & 1 & $0.066$ & 1 & ~ 		\\
		~ 										& $\left(2.2,0\right)$		& 1 & 0 & $0.066$ & 1 & NAND 	\\
		~ 										& ~ 		  								& 1 & 1 & $0.132$ & 0 & ~ 		\\
		~											& \multicolumn{6}{l}{\hrulefill}  										\\
		~ 										& ~ 		  								& 0 & 0 & $0.066$ & 1 & ~ 		\\
		~ 										& ~ 		  								& 0 & 1 & $0.132$ & 0 & ~		 	\\
		$\left(0,0.066\right)$& $\left(2.2,0.066\right)$& 1 & 0 & $0.132$ & 0 & NOR		\\
		~ 										& ~ 		  								& 1 & 1 & $0.198$ & 0 & ~			\\
		~											& \multicolumn{6}{l}{\hrulefill}  												\\
		~ 										& ~ 		  								& 0 & 0 & $0$ 		& 1 & ~ 		\\
		~ 										& ~ 		  								& 0 & 1 & $0.066$ & 0 & ~		 	\\
		~ 										& $\left(2.1,0\right)$		& 1 & 0 & $0.066$ & 0 & NOR			\\
		~ 										& ~ 		  								& 1 & 1 & $0.132$ & 0 & ~ 	 			
		\end{tabular}
		\end{ruledtabular}
	\end{center}
	\label{tab:ParameterCumTruthTable2}
\end{table}

\begin{figure}[ht]
      \includegraphics[scale = 0.325]{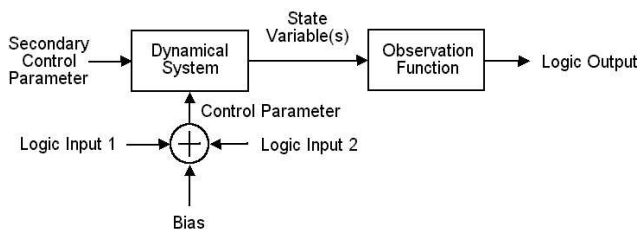}
\caption{Implementation of a logic-gate using a 2-dimensional map or a flow.  The transfer-function of the gate can be controlled using the secondary control parameter.}
\label{fig:scheme2}
\end{figure}

Variation in the critical value of the control parameter with the variation of a secondary control parameter can be seen in many 2-dimensional maps and can be used for controlling the gate's transfer-function as seen in this example.  In section \ref{sec:flows}, we shall see that such control of the transfer-function is also possible with continuous-time flows.  Fig. \ref{fig:scheme2} summarizes this scheme.

\section{Computation using Flows}	\label{sec:flows}
\begin{figure*}[ht]
	\includegraphics[scale = 0.27]{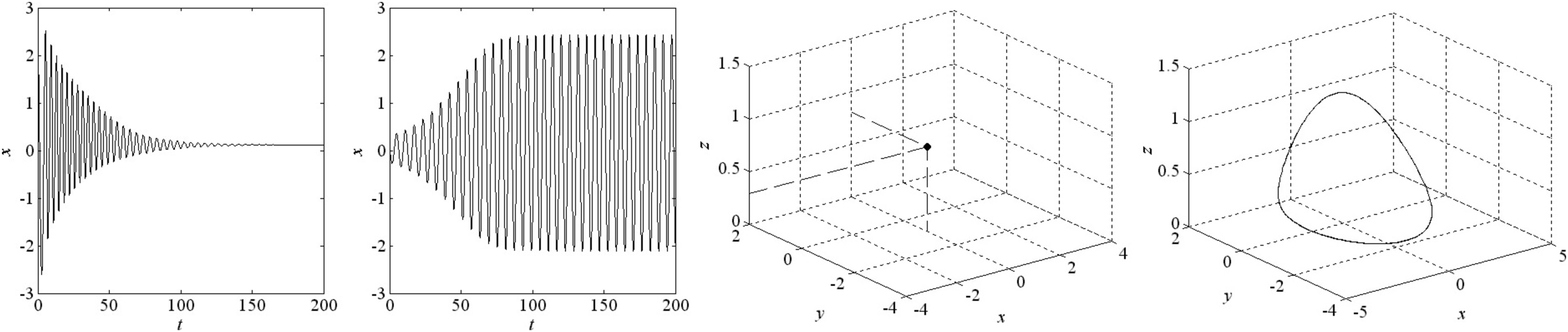}
  \begin{tabular}{cccc}
      \makebox[0.242\linewidth][c]{(a)} & \makebox[0.242\linewidth][c]{(b)} & \makebox[0.242\linewidth][c]{(c)} & \makebox[0.242\linewidth][c]{(d)}
  \end{tabular}
\caption{R\"ossler flow.  (a) $x\left(t\right)$ at $\upsilon = 0.5$.  (b) $x\left(t\right)$ at $\upsilon = 1.78$.  (c) Trajectory at $\upsilon = 0.5$.  (d) Trajectory at $\upsilon = 1.78$.}
\label{fig:timeseries}
\end{figure*}

Consider the R\"{o}ssler system \cite{peinke92, nagashima99}
\begin{equation}
  \left.
  \begin{array}{lcl}
    \dot{x} & = & -y -z \\
    \dot{y} & = & x + \sigma y \\
    \dot{z} & = & \tau + z\left(x - \upsilon\right)
  \end{array}
  \enspace \enspace \right\}
\end{equation}
where $\dot{x}, \dot{y}$ and $\dot{z}$ respectively denote the derivatives of the state-variables $x, y$ and $z$ with respective to time $t$.  Let $\sigma = 0.15$ and $\tau = 0.3$ .  Let $\upsilon$ be the control parameter.  The system bifurcates at $\upsilon \approx 1.14$ with fixed-point convergence for $\upsilon < 1.14$ and limit-cycle for $\upsilon > 1.14$.  Plots of state-variable $x(t)$ and stationary phase-space trajectories at $\upsilon = 0.5$ and  $\upsilon = 1.78$ can be seen in Fig. \ref{fig:timeseries}.

If a fixed-point denotes logic 0, a limit cycle denotes logic 1, ${\upsilon}' = 0.5, L_0=0$ and  $L_1 = 0.457$ we obtain an AND gate.  The bias value ${\upsilon}'=0.927$ yields an OR gate.

Several observation functions can be defined to differentiate between the signals of Fig.s \ref{fig:timeseries}(a) and \ref{fig:timeseries}(b).  For an electrical system, if $x\left(t\right)$ is a voltage signal, a full-wave rectifier followed by a low-pass filter will give a constant DC signal for the limit-cycle case of Fig. \ref{fig:timeseries}(b) while giving zero output for the fixed-point case of Fig. \ref{fig:timeseries}(a) and can be considered a realizable observation function.

As in the case of two-dimensional maps, a secondary control parameter can also be used to control the transfer function.  If $\tau$ is changed to $0.19$, the bifurcation from a fixed-point to limit-cycle takes place at $\upsilon \approx 0.72$.  Then, for $\left({\upsilon}' = 0.5, \sigma = 0.15, L_0 = 0, L_1 = 0.427\right)$ itself, we obtain an OR gate.  The configuration of the R\"ossler flow as a NAND/NOR gate is summarized in Table \ref{tab:ParameterCumTruthTable3}.
\begin{table}[ht]
\caption{R\"{o}ssler flow as an AND/OR Gate.  $Y=0$ for a fixed point, $Y=1$ for a limit-cycle.  The transfer function can be controlled by bias ${\upsilon}'$ or secondary control parameter $\tau$.}
		\begin{center} 
		\begin{ruledtabular}
		\begin{tabular}{lllllll}
			$\left(L_0,L_1\right)$ & $\left(\sigma,\tau,{\upsilon}'\right)$ & $\upsilon_a$ & $\upsilon_b$ & $\upsilon$ & $Y$ & Gate \\ \hline
		~ 										& ~ 		  											& 0 & 0 & $0.5$ 	& 0 & ~			\\
		~ 										& ~ 		  											& 0 & 1 & $0.927$ & 0 & ~ 		\\
		~ 										& $\left(0.15,0.3,0.5\right)$		& 1 & 0 & $0.927$ & 0 & AND 	\\
		~ 										& ~ 		  											& 1 & 1 & $1.354$ & 1 & ~ 		\\
		~											& \multicolumn{6}{l}{\hrulefill}  															\\
		~ 										& ~ 		  											& 0 & 0 & $0.927$ & 0 & ~ 		\\
		~ 										& ~ 		  											& 0 & 1 & $1.354$ & 1 & ~ 		\\
		$\left(0,0.427\right)$& $\left(0.15,0.3,0.927\right)$	& 1 & 0 & $1.354$ & 1 & OR 		\\
		~ 										& ~ 		  											& 1 & 1 & $1.781$ & 1 & ~ 		\\
		~											& \multicolumn{6}{l}{\hrulefill}  															\\
		~ 										& ~ 		  											& 0 & 0 & $0.5$ 	&	0 & ~ 			\\
		~ 										& ~ 		  											& 0 & 1 & $0.927$ & 1 & ~ 		\\
		~ 										& $\left(0.15,0.19,0.5\right)$	& 1 & 0 & $0.927$ & 1 & OR 		\\
		~ 										& ~ 		  											& 1 & 1 & $1.354$ & 1 & ~ 	
		\end{tabular}
		\end{ruledtabular}
	\end{center}
	\label{tab:ParameterCumTruthTable3}
\end{table}

\begin{figure}[ht]
\begin{center}
\includegraphics[scale = 0.325]{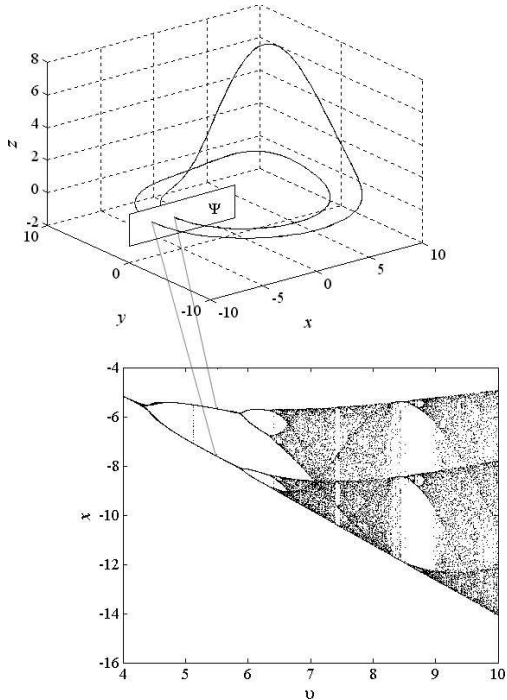}
\caption{The R\"{o}ssler Flow: One-sided Poincar\'{e} section for $\upsilon = 5.5$ (top) and bifurcation diagram (bottom).  The two points of the Poincar\'e section are shown projected on the corresponding points of the bifurcation diagram.}
\label{fig:bifurcation_rossler}
\end{center}
\end{figure}

The R\"{o}ssler system undergoes a series of bifurcations as $\upsilon$ is further increased.  Consider the one-sided Poincar\'{e} section \cite{parker87, kapitaniak98} formed by the set of points at which the stationary orbit intersects the $y=0$ plane as $y$ changes from positive to negative.  Let us denote this section plane by the letter $\Psi$.  The Poincar\'{e} section for $\upsilon = 5.5$ is shown in Fig. \ref{fig:bifurcation_rossler}(top) and consists of only two points.  Now, let $\upsilon$ be varied and at each value of $\upsilon$ the Poincar\'{e} section be determined.  The abscissa ($x$-coordinate) of each point in the Poincar\'e section, plotted against the corresponding value of $\upsilon$ gives the bifurcation diagram.  This is shown in Fig. \ref{fig:bifurcation_rossler}(bottom).  Any bifurcation that causes a qualitative change in the behavior of the system can be used to fashion the logic-gate around.

\section{Concluding Remarks}
It is not difficult to extend the proposed scheme to $m$-input logic gates provided the control parameter can be split into $m+1$ components corresponding to one bias and the $m$ logic inputs.

In the previously proposed scheme of using clipped chaotic dynamics for logical computations \cite{munakata02, murali03, murali03_2}, it is required to modify the dynamical system by inserting a threshold-controller in the feedback path.  The scheme proposed in this paper has no such requirement.  Further, the threshold controller required for clipped-chaos based computation is not easy to realize as a non-electrical system.  Therefore, the applications of such a scheme are likely to be restricted to electrical-realizations of dynamical systems.  However, the currently proposed scheme merely requires the system dynamics to be observed and a distinction to be made between the two possible states of the system.  Therefore, it is possible to employ dynamical systems of diverse physical nature - mechanical, chemical, optical, biological and fluid-dynamic systems - for information processing and computation.  

Since the NAND and NOR operations are \emph{functionally complete} \cite{kohavi96, tocci95}, it should be possible to implement any arbitrary switching function by interconnecting dynamical systems that realize these logic gates.  It is also interesting to note that the flip-flop, which is the fundamental digital memory element, can also be realized as an interconnection of NAND/NOR gates.

\bibliography{dynalogic}
\end{document}